\documentclass[twocolumn,twoside,slac]{revtex4}
\usepackage{graphicx}
\usepackage{fancyhdr}
\begin{document}

\title{S-Link to Gigabit Ethernet Adapter\\
New Frame Segmentation for LHCb Data Acquisition System}

\author{Joaquim E. Neves}
\affiliation{EP-Division, CERN, CH-1211 Geneva 23, Switzerland.\\
Escola de Engenharia da Universidade do Minho, P-4800-058 Guimar‹es, Portugal.}

\author{Richard  Jacobsson}
\author{Niko Neufeld}
\author{Beat Jost}
\affiliation{EP-Division, CERN, CH-1211 Geneva 23, Switzerland}

\begin{abstract}

Data Acquisition and Control Systems used in high energy physics experiments, such as those which will take place in the Large Hadron Collider (LHC) at CERN, require the specification of data formats and transmission protocols as well as the use of high speed links and interfaces.

In this context, a new Frame Segmentation process will be presented and discussed, based on data formats adopted by the LHCb experience for the interconnection of two standardized systems: S-link and Gigabit Ethernet. Simulation results of the transfer capacity of the proposed mechanism will be also reported, together with guidelines for its physical implementation.
\end{abstract}

\maketitle

\thispagestyle{fancy}

\section{INTRODUCTION}\label{s1}

The Large Hadron Collider beauty experiment for precise measurements of CP violation \cite{SLG1} and rare decays (LHCb) \cite{SLG2} requires High Speed Interconnect (HSI) Systems in order to transport the large amount of data generated by the detectors connected to the Front-end Motherboards (FEMB) to the storage devices connected to the Read-out Motherboards (ROMB). 

In the opposite direction, only a small volume of data generally has to be transferred between those parts for the control and management of the whole system.

The LHCb experiment adopted Gigabit Ethernet (GbE) \cite{SLG3} as the link technology from the output of the FEMB electronics boards to the input of the ROMB Sub-Farm Controllers (SFCs), and S-Link \cite{SLG4}, as a standard interface between the FEMB and the Data Acquisition (DAQ) system \cite{SLG2}.

\begin{widetext}

\begin{figure*}[h]
\centering
\includegraphics[width=140mm]{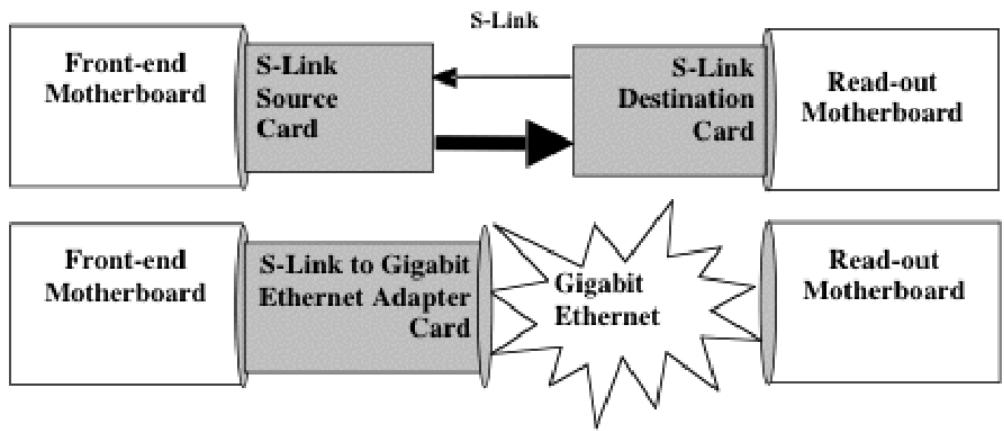}
\caption{- Architecture of the LHCb Data Acquisition System, based on S-Link to Gigabit Ethernet Adapter.} \label{f1}
\end{figure*}

\end{widetext}

The Architecture of the LHCb Data Acquisition, based both on full S-link transmission and on the S-Link to Gigabit Ethernet Adapter (SGbEA) System, is depicted in Figure  \ref{f1}. Since the Readout Network already supports the GbE, one S-Link card for the FEMB, based on GbE, was designed and built within the scope of the collaboration with the Atlas LHC experiment \cite{SLG5}. 

Next, in section \ref{s1}, the technical specification of the LHCb Data Acquisition System is discussed, while section \ref{s3} describes the new frame segmentation mechanism of the SGbEA. Simulation results for SGbEA throughput, using two different operation modes and data packet lengths, are presented in section \ref{s4}. The Section \ref{s5} deals with the physical implementation of the New Frame Segmentation and Section \ref{s6} summarizes the main conclusions.

\section{SYSTEM ARCHITECTURE}\label{s2}

The S-LINK specification defines a simple FIFO-like user interface at both ends of the transmission link which remains independent of the technology used to implement the physical link and which provides the transfer of event data and control words, error detection, optional flow control and test facilities \cite{SLG6}. The S-link specification also describes the interfaces between the FEMB and the Link Source Card (LSC) and between the Link Destination Card (LDC) and the ROMB, in either simplex or duplex version.

In the simplex version, since there is no communication path from the LDC to the LSC, the transmission is unidirectional, while, in the duplex version, the return channel between the LDC and the LSC allows the transmission of flow control commands from the ROMB to the FEMB. In both versions a single, high-density, 64-pin connector is used to connect the FEMB to the LSC and the LDC to the ROMB, allowing, per clock cycle, the transport of a 32 bit word at the frequency of 40 MHz, which corresponds to a throughput of 1.28 Gbit/s.

There is more than one solution for the implementation of the Front-End Multiplexers (FEMs) and the Readout Units (RUs). This is because, in the LHCb experiment, the main difference between these modules results from the fact that the RUs have to interface to the Readout Network, and hence must respect the GbE flow control protocol, which is not true for the links between FEMs and RUs \cite{SLG2}. 

\begin{widetext}

\begin{figure*}[b]
\centering
\includegraphics[width=145mm]{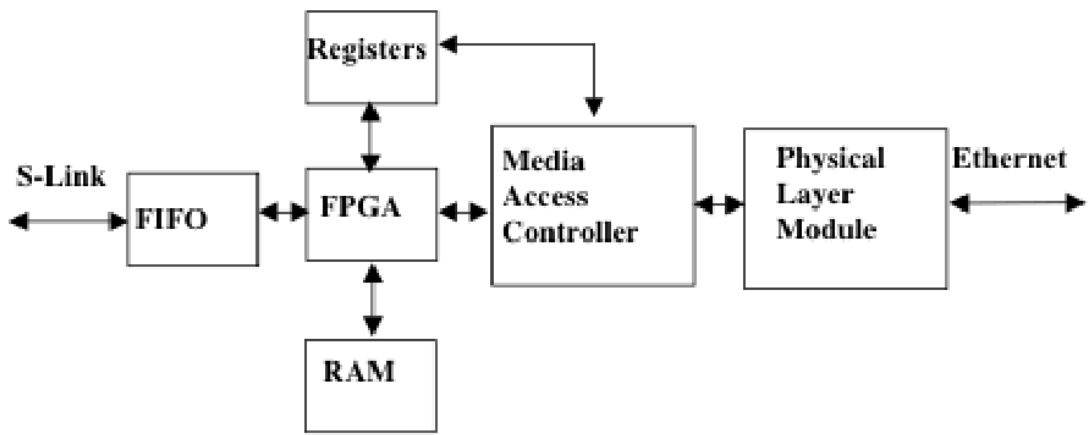}
\caption{- S-Link to Gigabit Ethernet Adapter Architecture.} \label{f2}
\end{figure*}

\end{widetext}

The SGbEA is a possible solution to interconnect FEMs and RUs, as is shown in Figure  \ref{f1}. This module has to implement the Ethernet functionalities of the Physical and Media Access Control (MAC) Layers, and the S-link specifications. The protocol conversion between S-link and GbE can be implemented within a FPGA, together with memories (FIFO, RAM and Registers), as can be seen in Figure \ref{f2}.

The SGbEA has to generate the Start Of Packet (SOP) and the End Of Packet (EOP) signals for the MAC device and to stop and restart the transmission when receiving watermark flags from the MAC or S-link FIFOs. Optionally, it can process the fragmentation of the S-link Packets on Ethernet frames, as is explained in the next section.

\section{NEW FRAME SEGMENTATION}\label{s3}  

Since within the communication links, between LHCb FEMs and RUs, the data can be transported with two different formats, the SGbEA has to support two operational modes: Short Packet Mode (SPM) and Long Packet Mode (LPM). 

In SPM, the length of the packets generated by the FEMB is variable, but always smaller than the maximum length of the Ethernet MAC frames. On the other hand, since the header of the LHCb data format already includes the header of the MAC Frames, no protocol conversion is necessary and the S-link packet is forwarded directly from the S-link FIFO to the MAC FIFO.

In LPM, the FEMB can generate data packets ranging in length from 52 Bytes to 32 Kbytes. 

In this operational mode the length of the S-link packet can be higher than the maximum length of the Ethernet MAC frames. For this reason, the header of each S-link packet, which contains a field with its length, has to be memorized in the SGbEA RAM, before calculating the number of fragments into which it must be split in order to be transmitted.  \newpage
In this mode, the SGbEA inserts, within each fragment header, the Type/length of the frame on the Ethernet header field, together with the number of fragments and the current fragment number of the S-link packet.

Optionally, in both operational modes, the Ethernet Sources and Destination Address can be generated by the FEMB (on a packet-by-packet basis), or can be previously inserted in the SGbEA Registers, by the control system and then transferred to each packet fragment header.

\section{SIMULATION RESULTS}\label{s4} 

A VHSIC Hardware Description Language (VHDL) model of the SGbEA module was developed for simulation purposes on the VisualHDL platform on a Computer-aided Engineering (CAE) system, together with an S-link packet generator.

Packets with different lengths and data formats were generated by this model at the S-link interface in either Long or Short Packet Mode, in order to be processed by the SGbEA and transmitted over the Gigabit Ethernet. 

In both cases the S-link overhead, introduced by the control words used for signaling the beginning and the end of the LHCb data packets, was reduced to a minimum: one word for the start and another one for the end of packet. In another interface, at the Ethernet side, in addition to the MAC overhead, a minimum inter frame gap of 12 bytes was also guaranteed.

\subsection{Short Packet Mode}  

Since in SPM there is no frame segmentation, the inclusion of a MAC frame header at the beginning of the packet is the only constraint imposed on the data format.

Figures \ref{f3} and \ref{f4} report some results for the SGbEA simulation in SPM. The maximum throughput for different packet lengths is present in Figure \ref{f3}, while Figure \ref{f4} shows the relationship between the effective throughput reached with those packets, and Gigabit Ethernet capacity.

As the figures show, due to the overhead of the Ethernet frames, when short packets with a size to the order of tens of bytes are transferred, the throughput is clearly lower than the transmission capacity, both in the S-link and in the GbE interfaces. On the other hand, for longer packets with over a hundred bytes, the transfer rate begins to be limited by the effective GbE capacity.

It is interesting to note that, for 64 byte packets, the reference value for packet length on several interfaces of the LHCb data acquisition system, the throughput is near 1.4 M Packet/s, allowing the use of SGBE on these interfaces with a trigger rate of about 1.1 MHz.

\begin{widetext}

\begin{figure*}[b]
\includegraphics[width=160mm]{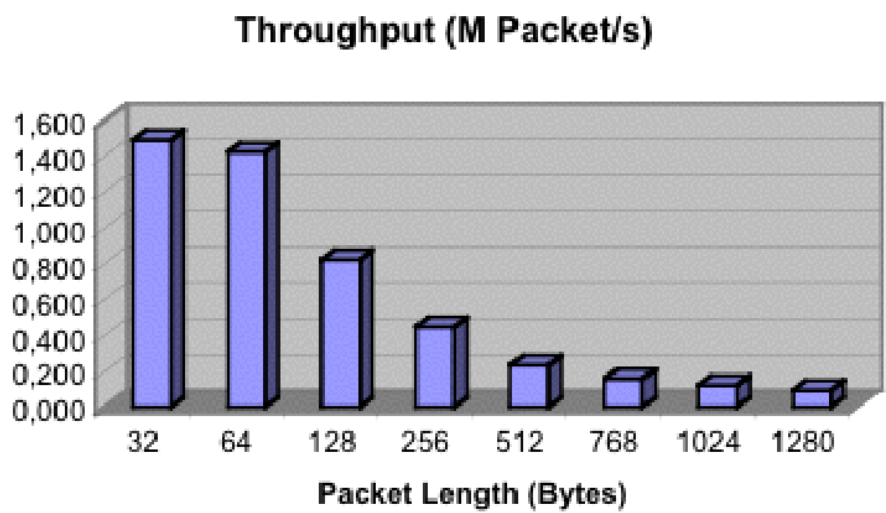}
\caption{- Maximum throughput achieved by the SGbEA with Short Packet Mode, for different packet length.} \label{f3}
\end{figure*}
\newpage
\begin{figure*}[t]
\includegraphics[width=150mm]{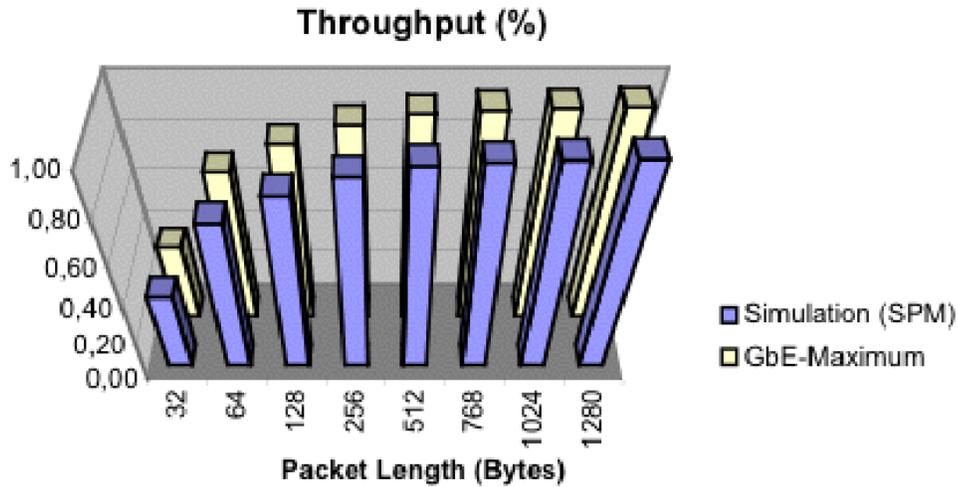}
\caption{- Maximum throughput achieved by Short Packet Mode and Gigabit Ethernet capacity for different packet length.} \label{f4}
\end{figure*}

\end{widetext}

\subsection{Long Packet Mode} 

Figures \ref{f5} and \ref{f6} present simulation results for Frame Segmentation in LPM. The maximum throughput of the system is shown in Figure \ref{f5}, while Figure \ref{f6} shows the relationship between the effective throughput achieved on the transmission of those packets and Gigabit Ethernet capacity.

In this mode, the LHCb data packets presented on the S-link interface are of variable length, as in SPM, but now the packet header has 52 bytes, including the header of the MAC frame.
\begin{widetext}

\begin{figure*}[b]
\includegraphics[width=150mm]{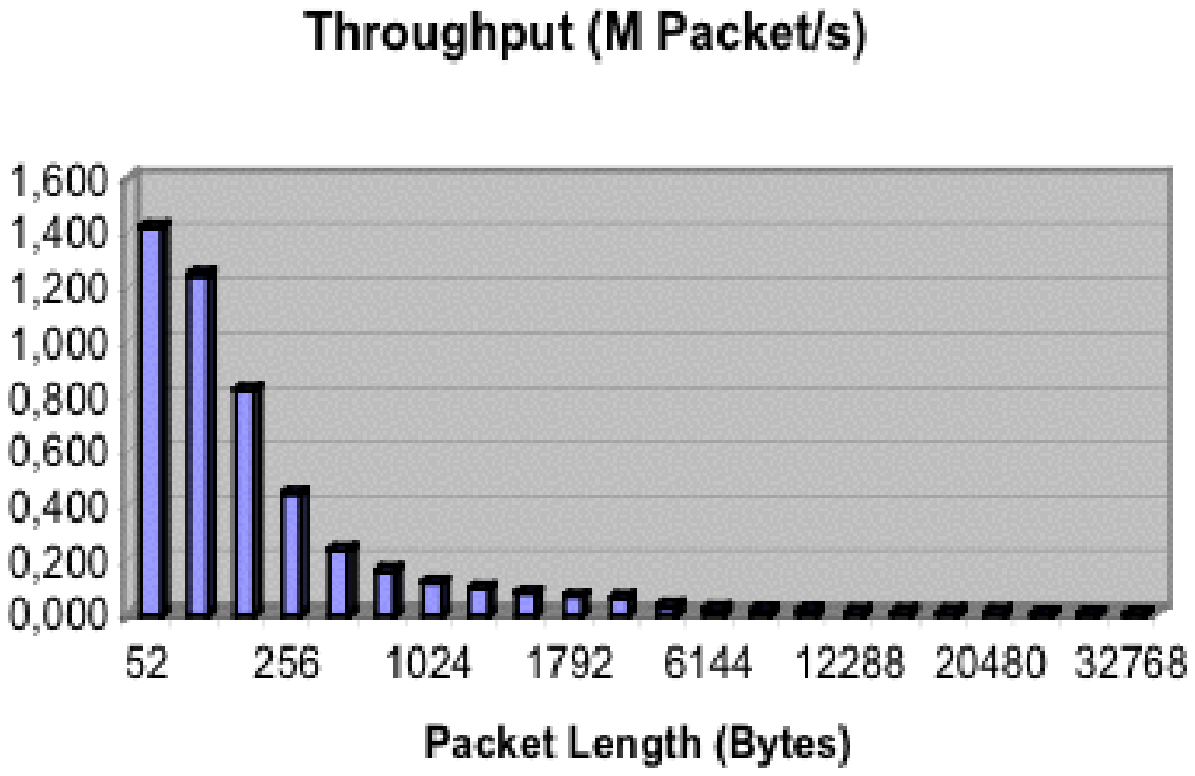}
\caption{- Maximum throughput achieved by SGbEA in Long Packet Mode, for different packet lengths.} \label{f5}
\end{figure*}
\newpage
\begin{figure*}[t]
\includegraphics[width=150mm]{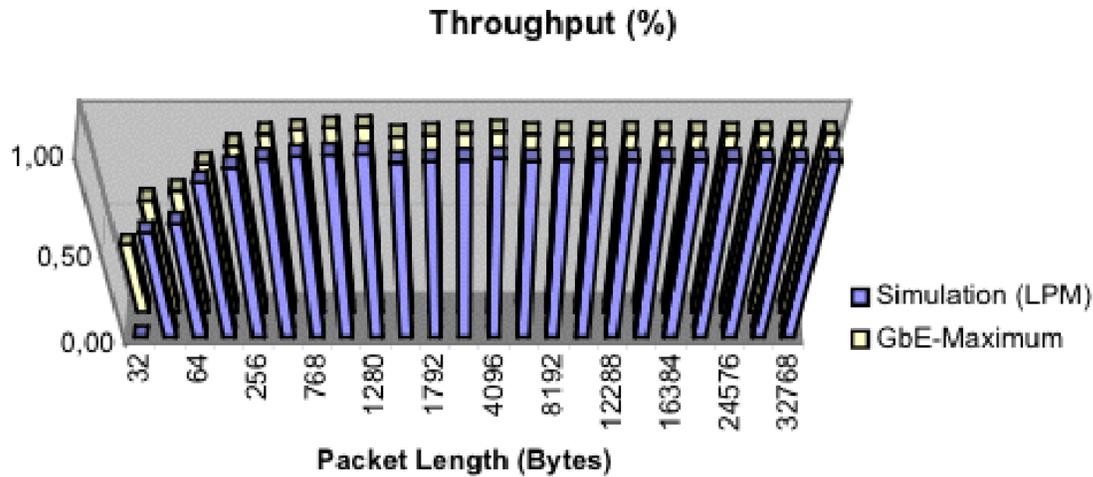}
\caption{- Maximum throughput achieved in Long Packet Mode and Gigabit Ethernet capacity for different packet lengths.} \label{f6}
\end{figure*}

\end{widetext}
As was expected, Figure 5 shows that the throughput achieved for packets of a length of less than 64 bytes is lower than that achieved in SPM, but is still higher than 1,2 M Packet/s, which means that the LPM can also be used with these packet lengths on the interfaces with a trigger rate of about 1,1 MHz. 

For longer packets, the decrease of the throughput is proportional to the increase of packet length, as the overhead introduced by the fragmentation process becomes negligible. As is depicted in Figure 6, the effective occupancy of the Ethernet payload increases with the length of the packet for short length packets - as was observed in the SPM simulations - and remains constant with longer packets.

\section{PHYSICAL IMPLEMENTATION}\label{s5}

The Frame Segmentation mechanism for LHCb Data Acquisition System described previously has been implemented within a FPGA on the S-Link to Gigabit Ethernet Adapter board itself, developed at the Argonne National Laboratory for Atlas LHC Experience. This prototyping board was implemented on a small daughter board over a PCI \footnote{PCI, Peripheral Component Interface.}  Mezzanine Card (PMC), according to the IEEE Common Mezzanine Card standard \cite{SLG7}.

A new version of the SGbEA has been specified for implementation in the near future. Since this new board will already be connected to the FEMB, the new Frame Segmentation can be fully tested, at the maximum S-link throughput, without the present constrains of the PCI Bus. 

The single port MAC Controller \cite{SLG8} could be replaced by a dual port \cite{SLG9} that supports the Gigabit Media Independent Interface (GMII) and the Ten-Bit Interfaces (TBI), allowing the Ethernet transmission over copper and/or fiber media. The Physical Layer module could also be implemented with a single dual port transceiver, either for copper or fiber media \cite{SLG10}. While one serial port is used in the copper interface, by connecting this port to a magnetic device \cite{SLG11}, the other is used in the fiber interface, by connecting this port to the optical module.

\section{CONCLUSIONS}\label{s6}

A new Frame Segmentation mechanism, with flow control capabilities, has been implemented over the S-Link to Gigabit Ethernet Adapter for the LHCb Data Acquisition System.

This mechanism is not only able to halt the transmission, in case of overflow, but can also optionally insert the Ethernet Source and Destination Addresses in the MAC frames.

As the simulation results show, this technique permits the transmission of the Short and/or Long Packets onto High Speed Interconnect Systems, which will be used in LHC experiments at CERN.

\end{document}